\documentstyle[11pt,paspconf,epsf]{article}

\begin{document}

\title{Particle Heating in Advection-Dominated Accretion
Flows}\affil{to appear in Proc. of ``The Workshop on High Energy
Processes in Accreting Black Holes,'' eds. J. Poutanen \& R. Svensson}

\author{Eliot Quataert}
\affil{Harvard-Smithsonian Center for Astrophysics, 60 Garden St., Cambridge, MA 02138; equataert@cfa.harvard.edu}



\begin{abstract}
I review particle heating by MHD turbulence in collisionless plasmas
appropriate to advection-dominated accretion flows.  These
considerations suggest that the preferential turbulent heating of
protons assumed by theoretical models is only achieved for relatively
subthermal magnetic fields. 
\end{abstract}




\section{Introduction}

A number of authors have argued that, at sub-Eddington accretion
rates, the gravitational potential energy released by turbulent
stresses in an accretion flow may be stored as thermal energy, rather
than radiated as in thin accretion disks (Ichimaru 1977; Rees et
al. 1982; Narayan \& Yi 1994, 1995; Abramowicz et al. 1995; see
Narayan, Mahadevan, \& Quataert 1998 and Kato, Fukue, \& Mineshige
1998 for reviews).  In this case, the gas heats up to nearly virial
temperatures and is assumed to adopt a two-temperature configuration,
with the protons significantly hotter than the radiating electrons.

There are two crucial microphysical issues relevant to these accretion
models, which are called either ion tori (Rees et al. 1982) or
advection-dominated accretion flows (ADAFs; Narayan \& Yi 1994).  The
first is the value of $\delta$, the fraction of the turbulent energy
in the plasma which heats the electrons (a fraction $1 - \delta$ heats
the protons).  The second is whether or not the dominant process for
exchanging energy between electrons and protons is Coulomb collisions.
ADAF models typically assume that $\delta \ll 1$ (preferential proton
heating) and that Coulomb collisions represent the only thermal
coupling.  In this case, the plasma is unable to cool because it is
nearly collisionless.  The protons, which by assumption receive the
turbulent energy, are (1) unable to radiate and (2) unable to transfer
their thermal energy to the electrons, the more efficient radiators.

The purpose of this review is to present some rather simple, but
general, theoretical considerations regarding particle heating in
ADAFs, which aim to address the value of $\delta$.  

\section{Physical Picture}

A fundamental assumption of my analysis is that the particle
distribution functions, while not necessarily strictly thermal, are
not pathologically nonthermal; that is, I assume that the plasma does
not reach an equilibrium configuration with, e.g., highly anisotropic
or bump-in-tail distribution functions.

For the simple considerations presented here, only two properties of
the plasma are important.  The first is the proton to electron
temperature ratio, $T_p/T_e$, taken to be $\gg 1$.  Note that it is
consistent to assume $T_p \gg T_e$ while simultaneously investigating
particle heating.  This is because the two temperature nature of the
plasma is in large part due to the efficiency with which relativistic
electrons cool; it is rather insensitive to $\delta$.  The second
property of importance is the magnetic field strength in the plasma,
parameterized by $\beta \equiv P_{\rm gas}/P_{\rm mag}$.  The quantity
$\beta$ must be $\ge 1$ for the gravitationally confined plasmas of
interest.

Like thin disks, ADAFs are turbulent magnetized plasmas.  On scales
much larger than the Larmor radius of the thermal protons ($\equiv
\rho$), we can treat the turbulence using MHD, i.e., as a
superposition of the magnetosonic and Alfv\'en waves.  Particles are
heated when the turbulence cascades to sufficiently small scales that
it is dissipated.  

\section{Collisionless Damping}

Let a wave in the plasma have wave vector $\bf k$ and frequency
$\omega$ and define $\perp$ and $\parallel$ to be directions
perpendicular and parallel to the local magnetic field.

In collisionless plasmas, the wave dissipation mechanisms of interest
are wave-particle resonances (molecular viscosity, thermal
conductivity, electrical resistivity, etc. are entirely unimportant).
In the next section, I will argue that the turbulence always has
frequencies well below the proton cyclotron frequency.  In this case,
resonance occurs when the wave's phase speed along the field line, $v
= \omega/k_\parallel$, equals $v_\parallel$, the velocity of a
particle along the field line.  A necessary (but not sufficient)
condition for strong damping is that $v$ be comparable to the thermal
speed of the particles, so that there are a large number of resonant
particles.

This resonance actually corresponds to two physically distinct
wave-particle interactions. In Landau damping (LD), particle
acceleration is due to the wave's longitudinal electric field
perturbation (i.e., the usual electrostatic force, $E_\parallel$).
For a wave damped solely by LD, $\delta \sim 1$ since the electrons
are preferentially heated by a factor of $\approx (m_e T^3_p/m_p
T_e^3)^{1/2}$, which is $\gg 1$ for $T_p \sim 10^2-10^3 T_e$.  In
transit-time damping (TTD), the magnetic analogue of LD (and the
collisionless analogue of Fermi acceleration; Achterberg 1981), the
interaction is between the particle's effective magnetic moment ($\mu
= m v^2_{\perp}/2 B$) and the wave's longitudinal magnetic field
perturbation, $B_\parallel$ (Stix 1992).  For a wave (with $k_\perp
\rho < 1$) damped solely by TTD, the protons are preferentially heated
by a factor of $\approx (m_p T_p/m_e T_e)^{1/2} \gg 1$; the
contribution to $\delta$ is therefore always $\ll 1$.  This is because
in plasmas with $T_p \gg T_e$, the protons have the larger magnetic
moment and so couple better to a wave's magnetic field perturbation.

%

\section{MHD Turbulence}

In a plasma with $\beta > 1$, one can (roughly speaking) decompose MHD
turbulence into three modes, the Alfv\'en wave and the fast and slow
magnetosonic modes.  The fast mode has the dispersion relation $\omega
\approx v_p |{\bf k}|$ ($v_p$ is the proton thermal speed) and is
essentially a sound wave.  The slow mode has $\omega \approx v_A
|k_\parallel|$ ($v_A \approx v_p \beta^{-1/2}$ is the Alfv\'en speed)
and, for $k_\perp \ne 0$, has a parallel magnetic field perturbation
($B_\parallel$).  The Alfv\'en wave also has $\omega \approx v_A
|k_\parallel|$, but is incompressible, with $E_\parallel = B_\parallel
= 0$ in the limit that the wavelength is long compared to the Larmor
radius of thermal protons ($|{\bf k}| \rho \ll 1$).

\subsection{Sound Waves (Fast Modes)}

Sound waves in an unmagnetized collisionless plasma are strongly
damped by LD.  For $T_p \gg T_e$, the dissipation time is formally
$\ll$ the mode period, implying that the modes are non-propagating.
These modes are unlikely to be efficiently excited since, quite
generally, strongly damped oscillators dissipate less energy than
weakly damped oscillators (consider, e.g., the analogous problem of
two resistors in parallel; the resistor with the smallest resistance
dissipates the most power). The same holds true for sound waves in a
$\beta > 1$ plasma, provided $k_\perp \sim k_\parallel$.  For $k_\perp
\gg k_\parallel$, i.e., nearly perpendicular sound waves, the parallel
phase speed is $v \approx v_p k_\perp/k_\parallel \gg v_p,v_e$ (since
$v_e \sim v_p$ by $T_p \gg T_e$).  Since the wave is highly
suprathermal, there are very few particles for the wave to resonate
with and it is consequently undamped.  These waves likely dissipate by
steepening and forming weak shock waves.

In what follows I will neglect quasi-perpendicular sound waves,
assuming that they are energetically unimportant.  The reason is that
I believe that the turbulence in ADAFs is intrinsically magnetic in
origin, being initiated by some combination of Balbus-Hawley and
convective instabilities.  The turbulent speed ($v_t$) is therefore
$\sim v_A < v_p$.  Moreover, I will argue that $v_A$ is likely to be
$\ll v_p$, in which case, if $v_t \sim v_A$, excitation of sound waves
will be strongly suppressed (subsonic instabilities do not easily
excite sound waves).

\subsection{Slow Modes}

In a $\beta \gg 1$ collisionless plasma, slow modes are strongly
damped and non-propagating for $k_\perp \ne 0$ (Foote \& Kulsrud
1979).  The reason is that, even in the limit of wavelengths long
compared to the Larmor radius of thermal protons ($|{\bf k}| \rho \ll
1$), slow modes have a parallel magnetic field perturbation
($B_\parallel \ne 0$) and are damped by TTD.  They are, as with sound
waves, unlikely to be excited.  For $k_\perp \rightarrow 0$, these
considerations fail, since $B_\parallel \rightarrow 0$.
Quasi-parallel slow modes are, however, likely to cascade in
$k_\perp$, develop $B_\parallel \ne 0$, and damp by TTD (heating the
protons).

\subsection{Alfv\'en Waves}

The Alfv\'en wave has a parallel phase speed $|v| = v_A$; in plasmas
appropriate to ADAFs, $v_A$ is comparable to the electron and proton
thermal speeds and so there are a large number of particles available
to resonate with the wave.  In the MHD limit, however, the Alfv\'en
wave is undamped by linear collisionless effects.  This is because
$E_\parallel = 0$ and $B_\parallel = 0$, i.e., there is no
electric/magnetic field perturbation associated with the wave which
can accelerate particles.

This lack of dissipation on large scales implies that understanding
heating of particles by Alfv\'enic turbulence requires understanding
how the energy cascades to small scales, where dissipation finally
becomes important.

\subsubsection{Alfv\'enic Turbulence}

The nature of Alfv\'enic turbulence in a collisionless, $\beta > 1$,
plasma is not fully understood. A plausible scenario is due to
Goldreich and Sridhar (1995; GS). They argue that Alfv\'enic
turbulence naturally evolves into a critically balanced state in which
the timescale for nonlinear effects to transfer energy from a
wavevector $\sim \bf k$ to a wavevector $\sim 2 \bf k$ ($\equiv$ the
cascade time, $T_c$) is comparable to the linear wave period at that
scale, $T = 2 \pi \omega^{-1}$; this determines how rapid the
dissipation must be to halt the cascade.  It also implies that the
cascade is highly anisotropic, with the energy cascading primarily
perpendicular to the local magnetic field; the parallel and
perpendicular sizes of a wave at any scale are correlated, with
$k_\parallel \sim k^{2/3}_{\perp} L^{-1/3} \ll k_\perp$, where $L$ is
the outer scale of the turbulence.

Quasi-perpendicular Alfv\'enic turbulence has the important property
that its frequencies are always much less than the proton cyclotron
frequency.  This is because the cascade occurs primarily in $k_\perp$
while the Alfv\'en frequency is $\propto k_\parallel$.  This justifies
my claim in \S3 that low frequency wave-particle resonances are of
primary importance.

\subsubsection{Alfv\'en Wave Damping}

The character of an Alfv\'en wave changes when $k_\perp \rho \sim 1$; at
this point the wave develops a non-zero parallel magnetic field
perturbation (in contrast to long wavelength perturbations, for which
$B_\parallel = 0$).  The waves can therefore be dissipated by TTD
(Quataert 1998, Gruzinov 1998, Quataert \& Gruzinov 1998; QG).

Figure 1 shows kinetic theory calculations of the dissipation of
nearly perpendicular Alfv\'en waves.  For $k_\perp \rho < 1$ (where
the waves can still be called Alfv\'en waves), the dissipation is
independent of $T_p/T_e$, but is a strong function of $\beta$.  The
electron temperature is unimportant since the electrons do not
participate significantly in the damping.  The dissipation is
sensitive to $\beta \approx (v_p/v_A)^2$, since this determines the
number of resonant protons and the protons' magnetic moment (which is
the effective coupling coefficient between the protons and the wave).

\begin{figure}
\begin{center} \leavevmode 
\hbox{%
\epsfxsize=5cm  
\epsfysize=5cm \epsffile{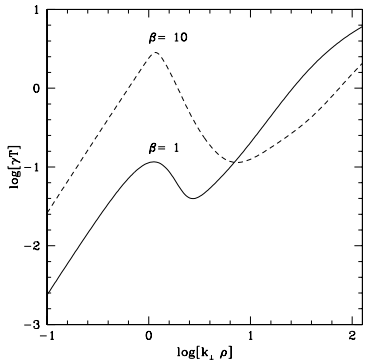}} 
\end{center} 
\caption{Dimensionless dissipation rate of an Alfv\'en wave ($k_\perp
\rho < 1$) or whistler wave ($k_\perp \rho > 1$) as a function of
perpendicular wavenumber.} \label{fig-1}
\end{figure}

For large $\beta \gg 1$, Alfv\'en waves are strongly damped for $k_\perp
\rho \sim 1$.  Most of the turbulent energy heats the protons, since
they are responsible for TTD (\S3).  For $\beta = 1$, however, the
maximal dissipation rate for an Alfv\'en wave in the GS cascade
(obtained at $k_\perp \rho = 1$) is $\gamma T \approx 0.1$ (see Figure
1).  Since the timescale for energy to cascade through the inertial
range is $T_c \approx T$, this suggests that, for plasmas with
equipartition magnetic fields ($\beta \approx 1$), very little of the
turbulent energy is dissipated on scales comparable to or greater than
$\rho$.  

Alfv\'en waves only exist for $k_\perp \rho < 1$.  For $k_\perp \rho >
1$, the same mode is called the whistler.  As emphasized by Gruzinov
(1998; see also QG), whistlers are unlikely to heat the protons.  For
$k_\perp \rho > 1$, but $k_\perp \gg k_\parallel$, whistlers have
$\omega \ll \Omega_p$.  Thus, in a mode period, a particle undergoes
many Larmor orbits.  Since the mode's perpendicular wavelength is
smaller than the proton Larmor radius ($k_\perp \rho > 1$), the
protons (but not the electrons) sample a rapidly varying
electro-magnetic field in the course of a Larmor orbit.  As a result,
they are ``frozen out'' and become dynamically unimportant; the
protons effectively no longer ``see'' the wave.  They therefore cannot
contribute to damping the whistler energy, which cascades to smaller
length scales until it is damped by the electrons (this is the origin
of the strong damping at $k_\perp \rho \gg 1$ in Figure 1).

\subsubsection{The Physics of $\delta$}

The above arguments suggest that the crucial physics determining the
electron heating rate in ADAFs is what fraction of the turbulent
energy is dissipated as Alfv\'en waves (heating the protons) and what
fraction cascades to small scales, becoming whistlers and heating the
electrons.  This depends on $\beta$, but not on $T_p/T_e$.
Unfortunately, estimates of the fraction of the turbulent energy that
becomes whistlers are highly uncertain, as they are exponentially
sensitive to the details of turbulence on the scale of the proton
Larmor radius, where both nonlinear and kinetic effects are important.
For example, we do not accurately know $T_c$, the nonlinear timescale
of the turbulence.  Nor is the calculation of the damping of the
turbulence that secure (because it is based on a theory of linear
Alfv\'en waves damped by thermal particles; both linearity and thermal
protons are crude approximations).

Parameterizing the uncertainty in the details of the turbulence, and
using numerical simulations of MHD turbulence as a guide, QG estimated
that the critical $\beta$ above which proton heating dominates (say
$\delta < 0.1$) is $\sim 10$, with an uncertainty of $\sim 10$!

\section{Summary and Future Work}

The physics of particle heating in a collisionless plasma suggests
that, in ADAFs and ion-tori, proton heating may be favored for
relatively subthermal magnetic fields, but not for strictly
equipartition ones ($\beta \equiv 1$).

In this paper, I have focused primarily on heating of particles by
small scale, incompressible, Alfv\'enic turbulence.  For $\beta \gg
1$, it should be quite accurate to neglect sound wave and slow mode
excitation.  As emphasized by Blackman (1998), however, one expects
the importance of compressibility to be an increasing function of
decreasing $\beta$.  For example, at $\beta = 1$, slow mode excitation
cannot be neglected (these modes are damped by TTD on relatively large
scales, heating the protons; e.g., Blackman 1998).  Compressibility
(and hence dissipation) of large scale Alfv\'en waves is also
potentially important.  My belief is that, while these effects are
undoubtedly present, it is unlikely that the majority of the turbulent
energy is dissipated on large scales; thus, to order of magnitude, the
analysis presented in this review is appropriate.  More quantitative
estimates are, however, clearly needed.

The unknown importance of magnetic reconnection, and its (presumed)
electron heating, remains a significant uncertainty for ADAF models
(Bisnovatyi-Kogan \& Lovelace 1997; see also QG).

\acknowledgments

It is a pleasure to acknowledge numerous discussions with Andrei
Gruzinov and Ramesh Narayan on the physics of ADAFs.  This work was
supported by an NSF Graduate Research Fellowship and by NSF Grant AST
9423209.

%

\end{document}